\documentclass[%
 reprint,
superscriptaddress,
 amsmath,amssymb,
 aps,
prd,
longbibliography,
]{revtex4-1}

\usepackage{aas_macros}
\usepackage{graphicx}
\usepackage{dcolumn}
\usepackage{bm}
\usepackage{hyperref}
\usepackage{orcidlink}

\usepackage{color}
\usepackage[mathlines]{lineno}
\usepackage{tabularx}
\usepackage{graphicx}
\usepackage{subfigure}
\usepackage[normalem]{ulem}
\usepackage[export]{adjustbox}
\usepackage{url}

\hypersetup{
    pdfnewwindow=true,    
    colorlinks=true,      
    linkcolor=blue,       
    citecolor=blue,       
    filecolor=blue,       
    urlcolor=blue         
}

\newcommand\stellarics{\texttt{StellarICs}}

\usepackage{natbib}

\begin{document}
\title{Anisotropic Photon and Electron Scattering without Ultrarelativistic Approximation}
\author{Anderson C.M. Lai\, \orcidlink{0000-0003-2741-4556}}
\email{cmlai@phy.cuhk.edu.hk}
\affiliation{Department of Physics, The Chinese University of Hong Kong, Shatin, New Territories, Hong Kong, China}

\author{Kenny C.Y. Ng, \orcidlink{0000-0001-8016-2170}}
\email{kcyng@cuhk.edu.hk}
\affiliation{Department of Physics, The Chinese University of Hong Kong, Shatin, New Territories, Hong Kong, China}

\date{November 28, 2022}

\begin{abstract}

Interactions between photons and electrons are ubiquitous in astrophysics. Photons can be down scattered (Compton scattering) or up scattered (inverse Compton scattering) by moving electrons. Inverse Compton scattering, in particular, is an essential process for the production of astrophysical gamma rays. Computations of inverse Compton emission typically adopts an isotropic or an ultrarelativistic assumption to simplify the calculation, which makes them unable to broadcast the formula to the whole phase space of source particles. In view of this, we develop a numerical scheme to compute the interactions between anisotropic photons and electrons without taking ultrarelativistic approximations. Compared to the ultrarelativistic limit, our exact results show major deviations when target photons are down scattered or when they possess energy comparable to source electrons.  We also consider two test cases of high-energy inverse Compton emission to validate our results in the ultrarelativistic limit. 
In general, our formalism can be applied to cases of anisotropic electron-photon scattering in various energy regimes, and for computing the polarizations of the scattered photons. 

\end{abstract}

\maketitle

\section{Introduction} \label{sec:Introduction}
\par
Interactions between electrons and photons are responsible for a variety of astrophysical phenomena. In particular, inverse Compton (IC) scattering---the up scattering of low-energy photons by high-energy electrons---is one of the main mechanisms for the production of astrophysical X-rays and gamma rays. For example, cosmic-ray electrons can IC scatter with the interstellar radiation field~\cite{Hunter:1997, Moskalenko:1999_puzzles, Strong:2000_diffuse_continuum_gamma_rays} and produce part of the Galactic diffuse gamma-ray emission~\cite{Thompson:1997_EGRET_paper, Strong:2004_diffuse_galactic_continuum, Strong:2004_new_determination_extragalactic_egret, Ackermann:2011_cocoon_of_freshly_accelerated_cosmic_rays_detected_by_fermi}. Cosmic-ray electrons can interact with solar photons and produce a gamma-ray halo around the Sun~\cite{Orlando:2007_gamma_rays_from_halos_around_sun, moskalenko:2006_solar_photons_modulation_and_neutrino_physics, Zhou:2017_TeV_solar_gamma_rays_from_cosmic_ray_interactions, orlando:2008_gamma_ray_from_solar_halo_and_disk_egret, orlando:2007_extened_IC_gamma_ray_from_sun, Fermi-LAT:2011, Kenny:2016_first_observation_solar_disk_with_fermiLAT, Orlando:2021_stellarics_sun_from_kev_to_tev, Tang:2018wqp, Linden:2020lvz, Linden:2018exo}. Gamma rays can be produced in astrophysical sources, such as pulsars, Blazars, and gamma-ray bursts, via external IC emission or synchrotron self-Compton~(SSC) emission~\cite{Chiang:1998_SSC1, Meszaros:1993_SSC2, Yuksel:2008rf, Linden:2017vvb, Sudoh:2019lav}. The up scattering of Cosmic Microwave Background (CMB) radiation is also responsible for the Sunyaev–Zeldovich (SZ) effect~\cite{Sunyaev:1970_SZ_effect}.  

\par
The IC emission formulation was studied in detail by Jones~\cite{Jones:1968}. Analytic expressions were obtained by considering isotropic distributions of electrons and photons for arbitrary electron energies. Jones also derived the photon spectrum from IC scattering in the ultrarelativistic limit, which was further developed by Blumenthal \& Gould (BG70 Hereafter)\cite{BG:1970} and Rybicki \& Lightman \cite{Rybicki:1986} by considering electrons with a power-law energy spectrum. These results have found numerous applications in high-energy astrophysics, such as the calculation of SSC emission associated with relativistic jets \cite{Inoue:2019jet1, Banik:2019jet2} and remnants from binary neutron star mergers \cite{Takami:2013_BNS}, etc.  While the ultrarelativistic results work well, the general formalism by Jones suffers from numerical instability over a broad range of photon and electron energies due to large number subtraction \cite{Belmont:2009_numerical_Compton}.

\par 
Substantial efforts have been devoted to mitigating the numerical instability in Jones' expression, including reformulations of or corrections to Jones' formula ~\cite{NP94_general_istropic_IC2, Peer:2004_correction}, as well as interpolation from ultrarelativistic limit \cite{Coppi:1990_midly-relativistic_interpolation}. These improvements have been applied to numerical calculations of radiative processes \cite{Peer:2005_application_general_isotropic_IC1, Vurm:2008_application_general_isotropic_IC2} and CMB spectral distortions \cite{Sarkar:2019_kinematic_regiems}. Nevertheless, these works only considered isotropic scattering, and usually focused on specific kinematic regimes and target energy distributions. (E.g., ultrarelativistic electrons or thermal photons/electrons.)

\par
For the more general cases of anisotropic photon-electron scatterings, Aharonian \& Atoyan~\cite{Aharonian:1981} derived the differential cross section for the scattering between anisotropic photons and isotropic electrons, and was applied in \cite{Chen:2011_application_anisotrpic_IC1, Murase:2010_application_anisotropic_IC2}.  These were also later rederived by \cite{NP93:anisotropic_IC} and \cite{Brunetti:anisotropic_IC}. 
In addition, Poutanen \& Vurm~\cite{Poutanen:2010_weakly_anisotropic_electrons} considered electrons in the weak anisotropic approximation scattering with isotropic photons. Kelner et. al.~\cite{Kelner:2014_anisotropic_electrons} studied anisotropic electrons scattering with isotropic photons but took the ultrarelativistic limit. Alternatively, a Monte Carlo approach was proposed to tackle the anisotropic scattering \cite{Molnar:1999_MC_anisotropic_IC}.  
The scattering between anisotropic photon and isotropic electrons, in the ultrarelativistic limit, was also considered by Moskalenko \& Strong (MS hereafter)~\cite{Moskalenko:2000_anisotropic_IC_in_galaxy}, which is often applied in computing the IC emission from cosmic-ray interactions in the galaxy~\cite{Strong:2000_diffuse_continuum_gamma_rays} and in the solar system~\cite{moskalenko:2006_solar_photons_modulation_and_neutrino_physics, Orlando:2007_gamma_rays_from_halos_around_sun}.  However, these analytic expressions were sometimes found to be numerically unstable (\cite{Vurm:2008_application_general_isotropic_IC2, Belmont:2009_numerical_Compton}). A general formalism for anisotropic electron-photon scatterings that is applicable in all kinematic regimes remains unavailable. 

\par
In this paper, we present a numerical integration approach that solves the general anisotropic IC scattering problem and calculates the resulting photon spectrum. In section~\ref{sec:Formulation}, we describe the formalism on how we handle the kinematic constraints arising from the differential cross section. In section~\ref{sec:Results}, we validate our calculations using solar IC and SSC emission as test cases. We show that our results are numerically stable and correctly converge in the ultrarelativistic limit; we also discuss the cases where the exact calculation deviate from the ultrarelativistic limit. We conclude and discuss the outlook of this work in section~\ref{sec:Conclusions}.

\section{Formulation} \label{sec:Formulation}

\subsection{Master Equation for IC intensity}
\par
\begin{figure}[t!]
    \centering
    \includegraphics[width=\columnwidth]{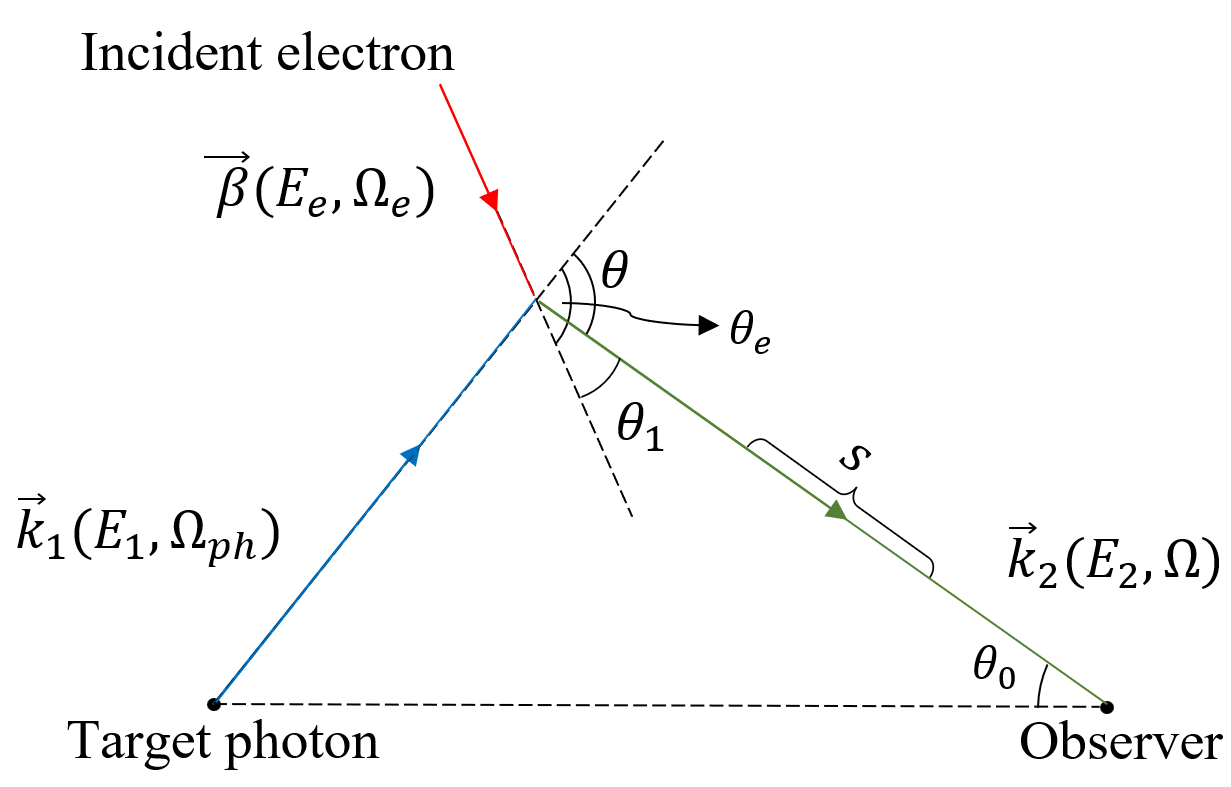}
    \caption{A schematic diagram that depicts the IC scattering for the case of a point source of target photons. $\vec{k}_1$, $\vec{k}_2$, and $\vec{\beta}$ are the directions of target photon, scattered photon and source electron. See the text for a detailed description of the variables. Note that in general $\vec{k}_1$, $\vec{k}_2$, and $\vec{\beta}$ do not lie in the same plane, see Fig.~\ref{fig:3d_scat_geom}.}
    \label{fig:2d_scat_geom}
\end{figure}

The master equation for line-of-sight (LOS) IC spectral intensity is given by:
\begin{align}\label{eq:master_equation}
    \frac{dI}{dE_2} (\Omega_o, E_2) = \frac{c}{4\pi}\int ds\int n_e dE_e \int n_{ph} dE_1 \left<\frac{d^2\sigma}{d\Omega dE_2}\right>,
\end{align}
where $\Omega_o(\theta_o, \phi_o)$ is the LOS direction w.r.t the observer, $s$ is the LOS distance, $E_e$, $E_1$, $E_2$ are the electron, the target photon, and the scattered photon energies, respectively. $n_e(E_e)$ is the differential number density of source electron flux, $n_{ph}(E_1)$ is the differential number density of target photon field. The angular averaged differential cross section is given by
\begin{align}\label{eq:ang_avg_cross_section}
    \left<\frac{d^2\sigma}{d\Omega dE_2}\right> = \int d\Omega_{ph} f_{ph} \int d\Omega_e f_{e} \frac{d^2\sigma}{d\Omega dE_2} (1 - \beta\cos\theta_e),
\end{align}
where $\Omega(\theta, \phi)$, $\Omega_{ph}(\theta_{ph}, \phi_{ph})$, and $\Omega_{e}(\theta_e, \phi_e)$ are the directions of the scattered photon, target photons, and source electrons, respectively. $f_{ph}(E_{ph}, \Omega_{ph})$ and $f_e(E_e, \Omega_{e})$ represent normalized angular distributions of target photons and source electrons with $\int f d\Omega =1$. $\beta$ is the speed of incident electrons and the factor $(1 - \beta\cos\theta_e)$ comes in as the correction factor on electron flux/photon density for non-parallel target photon and source electron.  We will use the simplified expression:
\begin{align}\label{eq:simplify_cross_section}
    \frac{d^2\sigma'}{d\Omega dE_2} = \frac{d^2\sigma}{d\Omega dE_2} (1 - \beta\cos\theta_e) 
\end{align}
hereafter.
The definitions of the variables are schematically shown in Fig.~\ref{fig:2d_scat_geom}. 

\subsection{Assumptions}
\par
In general, the differential cross section Eq.~\ref{eq:ang_avg_cross_section} depends on many variables. To simplify, we assume a unidirectional target photon along direction $\Omega'_{ph}$ from a point source, we also assume an isotropic distribution of source electron flux. That is:
\begin{equation}\label{eq:fe_fph}
    \begin{split}
        f_{ph} &= \delta({\Omega_{ph} - \Omega'_{ph}})\, , \\ 
        f_e &= \frac{1}{4\pi} \, .
    \end{split}
\end{equation}
Eq.~\ref{eq:ang_avg_cross_section} then becomes:
\begin{align}\label{eq:simplify_ang_avg_cross_section}
    \left<\frac{d^2\sigma}{d\Omega dE_2}\right> = \frac{1}{4\pi}\int d\Omega_e \frac{d^2\sigma'}{d\Omega dE_2}(E_1, E_2, E_e, \Omega, \Omega_e) ,
\end{align}
where $\Omega_{ph}$ is dropped by the delta function and $\Omega_e$ is now defined w.r.t to the direction of target photon.
The explicit form of the differential cross section in Eq.~\ref{eq:simplify_cross_section} is given by~\cite{jauch:2012theory}:
\begin{align}\label{eq:explicit_cross_section}
    \frac{d^2\sigma'}{d\Omega dE_2} = \frac{r_e^2 \bar{X}}{2\gamma^2(1 - \beta\cos\theta_e)}\frac{E_2^2}{E_1^2}\delta(E_2 - \bar{E}_2) ,
\end{align}
where $r_e$ is the classical electron radius, $\gamma$ is the electron Lorentz factor and
\begin{align}\label{eq:Xbar}
    \bar{X} = \left[\frac{\kappa_1}{\kappa_2} + \frac{\kappa_2}{\kappa_1} + 2 m_e^2\left(\frac{1}{\kappa_1} - \frac{1}{\kappa_2}\right) +  m_e^4 \left(\frac{1}{\kappa_1} - \frac{1}{\kappa_2}\right)^2\right] ,
\end{align}
where $\kappa_1 = p \cdot k_1 = p_0 k_0 - \vec{p}\cdot\vec{k}_1$ and $\kappa_2 = p \cdot k_2$ are the dot products of four momentum of incident electron and incident/scattered photon, the subscript 0 denotes the time component. $\bar{E}_2$ can be found from conservation of 4-momentum:
\begin{align}\label{eq:explicit_E2}
    \bar{E}_2 = \frac{E_1(1 - \beta\cos\theta_e)}{\frac{E_1}{E_e}(1 - \cos\theta) + 1 - \beta\cos\theta_1} ,
\end{align}
where $\theta_1$ is the angle between source electron and scattered photon,
\begin{align}\label{eq:cos_theta1}
    \cos\theta_1 = \cos\theta\cos\theta_e + \sin\theta\sin\theta_e\cos(\phi_e - \phi) .
\end{align}

We note that although Eq.~\ref{eq:cos_theta1} depends on 4 angular variables $\phi_e, \theta_e, \phi, \theta$, the two polar angles only come in as their difference $\phi_e - \phi$. This implies a polar symmetry on $\phi$ as we integrate $\phi_e$. In other words, Eq.~\ref{eq:explicit_cross_section} does not depend on the $\phi$, and we chose $\phi = 0$ hereafter.

\par
In the limit $\beta \ll 1$, Eq.~\ref{eq:explicit_E2} reduces to the Compton scattering formula and $\kappa_1, \kappa_2$ become $m_e E_1$ and $m_e E_2$. Putting these expressions back in Eq.~\ref{eq:explicit_cross_section} yields the Klein-Nishina (KN) differential cross section in electron-rest frame (ERF):
\begin{equation}\label{eq:KN_ERF_corss_section}
     \frac{d^2\sigma'_{\text{KN}}}{d\Omega dE_2} = \frac{r_e^2}{2}\frac{E_2^2}{E_1^2}\left(\frac{E_2}{E_1} + \frac{E_1}{E_2} - \sin^2\theta \right)\delta(E_2 - \bar{E}_2).
\end{equation}

\begin{figure}[t!]
    \includegraphics[width=\columnwidth]{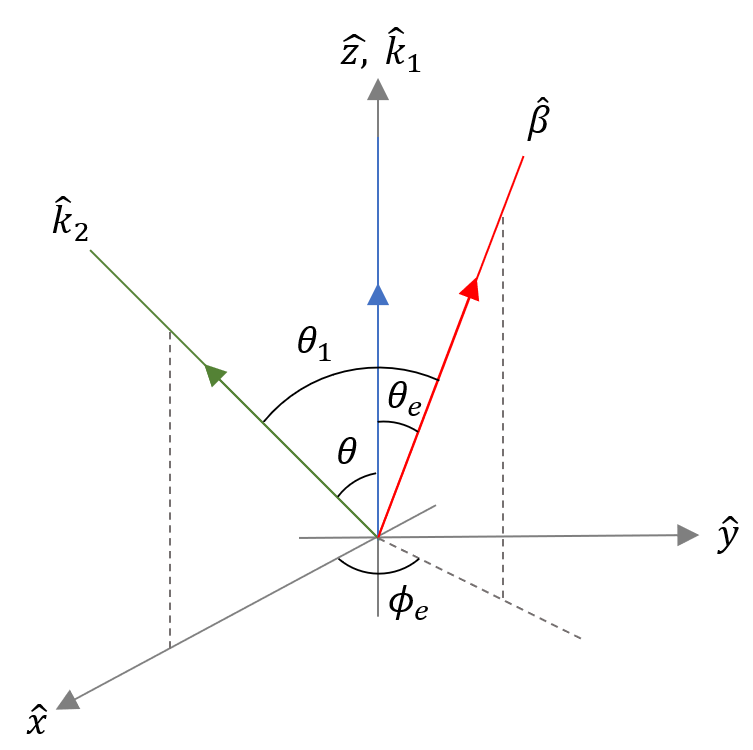}
    \caption{A diagram for the IC scattering between a photon and an electron. With the assumptions made in Eq.~\ref{eq:fe_fph}, we can choose to align the target photon direction with $+z-$axis and set $x-z$ plane as the scattering plane.}
    \label{fig:3d_scat_geom}
\end{figure}

\subsection{Reduction to MS result}
\par
As mentioned in the introduction, MS~\cite{Moskalenko:2000_anisotropic_IC_in_galaxy} adopted an ultrarelativistic assumption, in particular, $\gamma \gg 1$ and $\theta_1 = 0$. The latter implies that scattered photons are unidirectional along the direction of electron, or $\Omega = \Omega_e$. To illustrate the effect of such approximation, we first rewrite the delta function in Eq.~\ref{eq:explicit_cross_section} as:
\begin{align}\label{eq:delta_no_approx}
    \delta(E_2 - \bar{E}_2) = \frac{\delta(\Omega_e - \Omega_{e, sol})}{\left|\frac{d\bar{E}_2}{d\Omega_e}\right|},
\end{align}
where $\Omega_{e, sol}$ is the solution to the condition of $E_2 = \bar{E}_2(\Omega_{e, sol}) $. The unidirectional approximation is then equivalent to:
\begin{align}\label{eq:delta_approx}
    \delta(\Omega_e - \Omega_{e, sol}) &= \delta(\phi_e)~\delta(\cos\theta_e - \cos\theta)
\end{align}
\par
With this simplification, integration in Eq.~\ref{eq:simplify_ang_avg_cross_section} implies removing the $\Omega_e$ dependence and evaluate the denominator in Eq.~\ref{eq:delta_no_approx} at $\Omega_e = \Omega$. By restoring a general source photon distribution $f_{ph}$ and integrating over $\Omega_{ph}$, it can be shown that the whole expression reduces to MS's Eq.~(8)\cite{Moskalenko:2000_anisotropic_IC_in_galaxy}:
\begin{eqnarray}\label{eq:MS_equation}
        &\left<\frac{d\sigma}{dE_2}\right> = \frac{\pi r_e^2}{E_1(\gamma - E_2)^2}\int d\Omega_{ph}f_{ph}\left[2- 2\frac{E_2}{\gamma}\left(\frac{1}{E'_1} + 2\right)\right. \nonumber\\
        &\left.+ \frac{E_2^2}{\gamma^2}\left(\frac{1}{{E'}_1^2} + 2\frac{1}{E'_1} + 3\right)- \frac{E_2^3}{\gamma^3}\right].
\end{eqnarray}

\subsection{General treatment without unidirectional approximation}
\par
Instead of doing the unidirectional approximation in the last section, we look for the exact solution of $\cos\theta_{e, sol}$ in terms of $E_1, E_2, E_e, \theta$ and $\phi_e$. To begin with, we write the complete form of Eq.~\ref{eq:simplify_ang_avg_cross_section} using Eq.~\ref{eq:explicit_cross_section},~\ref{eq:Xbar} and~\ref{eq:delta_no_approx}: 
\begin{eqnarray}\label{eq:explicit_averaged_cross_section}
        &&\left<\frac{d^2\sigma}{d\Omega dE_2}\right>  \\
        &=& \frac{r^2_e}{8\pi\gamma^2}\int d\Omega_e \frac{\bar{X}E^2_2}{(1 - \beta\cos\theta_e)E^2_1}  \frac{\delta(\cos\theta_e - \cos\theta_{e, sol})}{\left|\frac{d\bar{E}_2}{d\cos\theta_e}\right|_{\cos\theta_{e, sol}}} \nonumber \\
        &=& \frac{r^2_e}{8\pi\gamma^2}\int^{2\pi}_0d\phi_e\sum_{\cos\bar{\theta}_{e, sol}}\left[\frac{\bar{X}E^2_2}{(1 - \beta\cos\theta_e)E^2_1\left|\frac{d\bar{E}_2}{d\cos\theta_e}\right|}\right] \nonumber \, ,
\end{eqnarray}
where
\begin{eqnarray}\label{eq:dE2_dcos_thetae}
        &&\frac{d\bar{E}_2}{d\cos\theta_e} \\
        &=& \frac{\beta \bar{E}_2}{1 - \beta\cos\theta_e}\left[\frac{ \bar{E}_2}{E_1}(\cos\theta - \sin\theta\cos\phi_e\cot\theta_e) - 1\right] \nonumber .
\end{eqnarray}

\begin{figure*}[t!]
    \centering
    \begin{subfigure}
        \centering
        \includegraphics[width=\columnwidth]{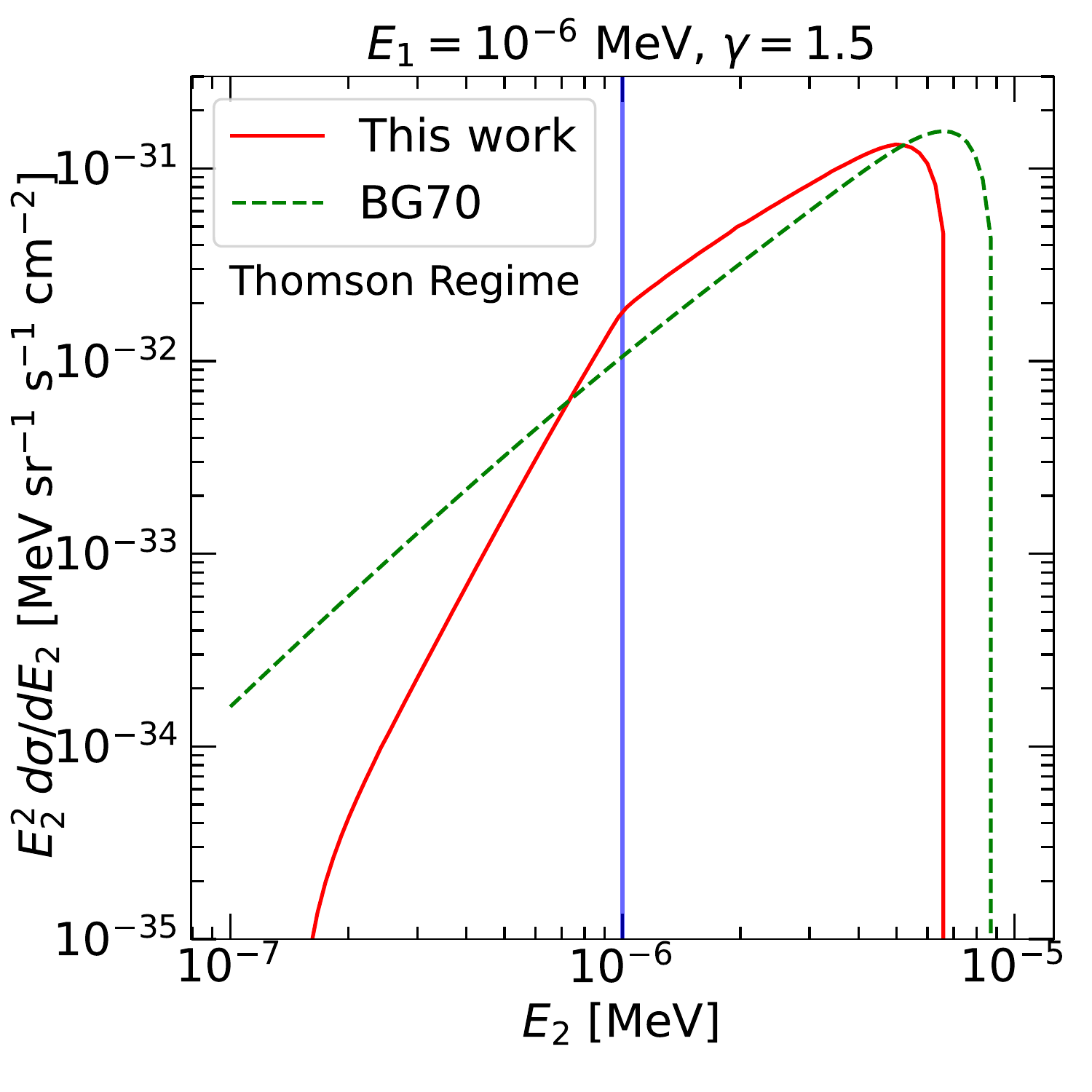}
        \label{fig:Thomson_small_gamma}
    \end{subfigure}
    \begin{subfigure}
        \centering
        \includegraphics[width=\columnwidth]{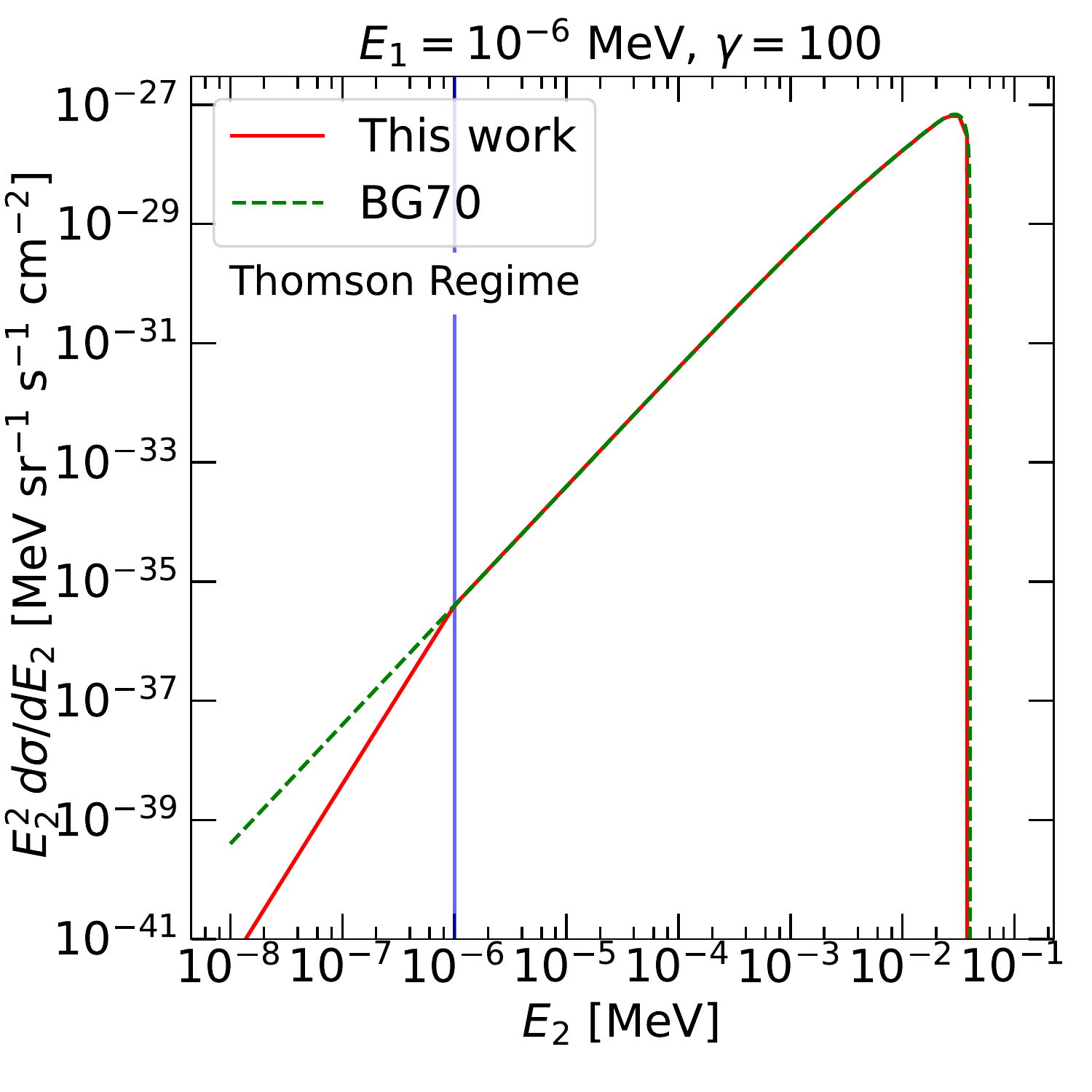}
        \label{fig:Thomson_large_gamma}
    \end{subfigure}
    \caption{The differential cross sections $E_2^2 \:d\sigma/dE_2$ against scattered photon energy $E_2$ in the Thomson regime $E_1 \ll E_e$ (Sec.~\ref{sec:Eph_ll_Ee}). Left: target photon energy $E_1 = 10^{-6}$ MeV, source electron Lorentz factor $\gamma = 1.5$. BG70 refers to the Eq. (2.48) in \cite{BG:1970}, the blue line marks the value of $E_1$. See paragraph for more details; Right: target photon energy $E_1 = 10^{-6}$ MeV, source electron Lorentz factor $\gamma = 100$.}
    \label{fig:Thomson_single_photon_electron}
\end{figure*}
\par

\par 
The problem of finding the differential cross section then reduces to summing and finding the possible solutions $\cos\theta_{e, sol}$, and then numerically integrate over $\phi_{e}$. It can be achieved by putting Eq.~\ref{eq:explicit_E2} into a quadratic equation of $\tan(\theta_e/2)$:
\begin{align}\label{eq:solve_thetae}
    (C - A)\tan^2\left(\frac{\theta_e}{2}\right) + 2B\tan\left(\frac{\theta_e}{2}\right) + (C + A) = 0 ,
\end{align}
where $A, B$ and $C$ are given by:
\begin{align}\label{eq:ABC}
    \begin{split}
        A &= \beta(E_1 - E_2 \cos\theta)\, ,\\
        B &= -E_2\beta\sin\theta\cos\phi_e\, ,\\
        C &= \frac{E_2E_1(1 - \cos\theta)}{E_e} + E_2 - E_1 .
    \end{split}
\end{align}
Hence, the two solutions to the equation
\begin{equation}\label{eq:theta_e_equation}
    \tan(\theta_e/2) = \frac{-B \pm \sqrt{A^2 + B^2 - C^2}}{C - A}
\end{equation}
correspond to two possible IC scattering geometry for a given set of $(E_2, E_1, E_e, \theta, \phi_e)$. Although both solutions lead to physical scattering geometries, only positive solutions are retained. That is because a negative $\tan(\theta_e/2)$ can be mapped to a positive one $\theta_e \rightarrow \theta_e + \pi$ together with $\phi_e \rightarrow \phi_e + \pi$. Therefore, it represents a duplicated a solution at another $\phi_e$. The relation between a positive determinant $A^2 + B^2 - C^2$ and the physical limit of $E_2$ is explained in Appendix~\ref{app:positive_determinant}.

\section{Results}\label{sec:Results}
\subsection{Scattering between isotropic photon and isotropic electrons}\label{sec:monoenergetic_electron_photon}

In this section, we compare the results obtained with our general formalism to that from BG70~\cite{BG:1970}, which was obtained using the high-$\gamma$ approximation and is used frequently in the IC literature.  

BG70 also assumed isotropic photon and electron distributions.  To match that, we evaluate the differential cross section by averaging over the scattering angle of Eq.~\ref{eq:explicit_averaged_cross_section}.  Although Eq.~\ref{eq:explicit_averaged_cross_section} is derived from isotropic source electron and unidirectional target photon (Eq.~\ref{eq:fe_fph}), averaging over the scattering angle is equivalent to averaging over the incident photon directions, and thus corresponds to an isotropic source photon distribution.  

We note that in the context of thermal SZ effect, Sarkar et. al.~\cite{Sarkar:2019_kinematic_regiems} also considered general isotropic scatterings. They defined the kinematic regimes by comparing the source photon energy $E_1$ and the electron energy $E_e$.  This is different from us, as we consider mainly the energies in the electron-rest frame (ERF). Below we refer to primed variables as ERF quantities (e.g., $E_{1}'$) and unprimed variables as the observer frame quantities (e.g., $E_{1}$).

\subsubsection{Thomson regime ($E_1' \ll m_e$)}\label{sec:Eph_ll_Ee}
\par
When the ERF target photon energy $E_1' = \gamma E_1(1 - \beta\cos\theta_e)$ is much smaller than the electron rest mass, the scattering between target photons and electrons falls into the Thomson regime regardless of the value of $\gamma$. In the Thomson limit, photon energies are the same before and after scattering in the ERF, $E_1' \approx E_2'$. In the ultrarelativistic limit ($\gamma \gg 1$), the maximum cutoff of the IC scattered photon energy in the observer frame, $E_{2, \rm max}$, is well approximated by $4\gamma^2 E_1$, which is the limit used by BG70.

\begin{figure*}[t!]
    \centering
    \begin{subfigure}
        \centering
        \includegraphics[width=\columnwidth]{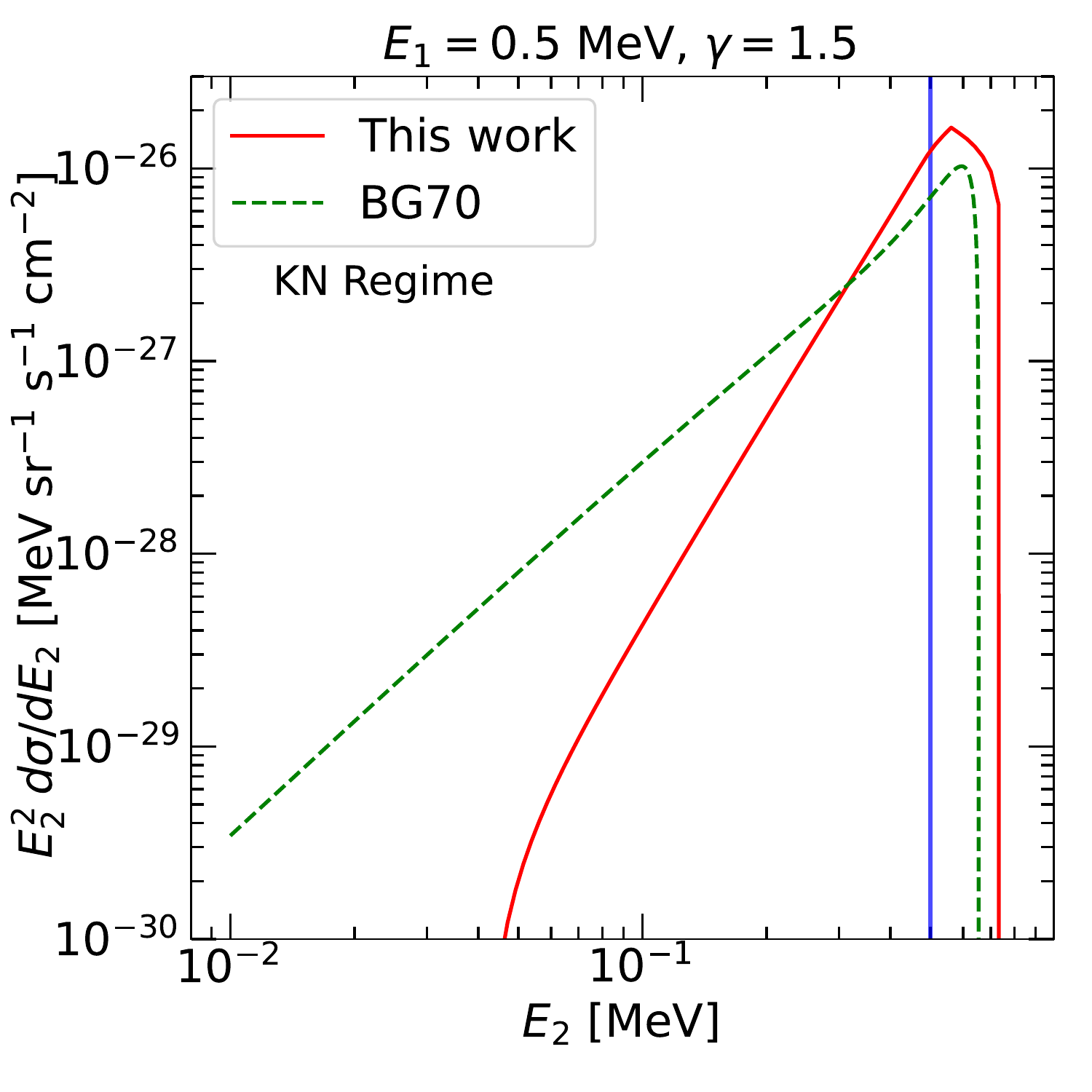}
        \label{fig:soft_KN_small_gamma}
    \end{subfigure}
    \begin{subfigure}
        \centering
        \includegraphics[width=\columnwidth]{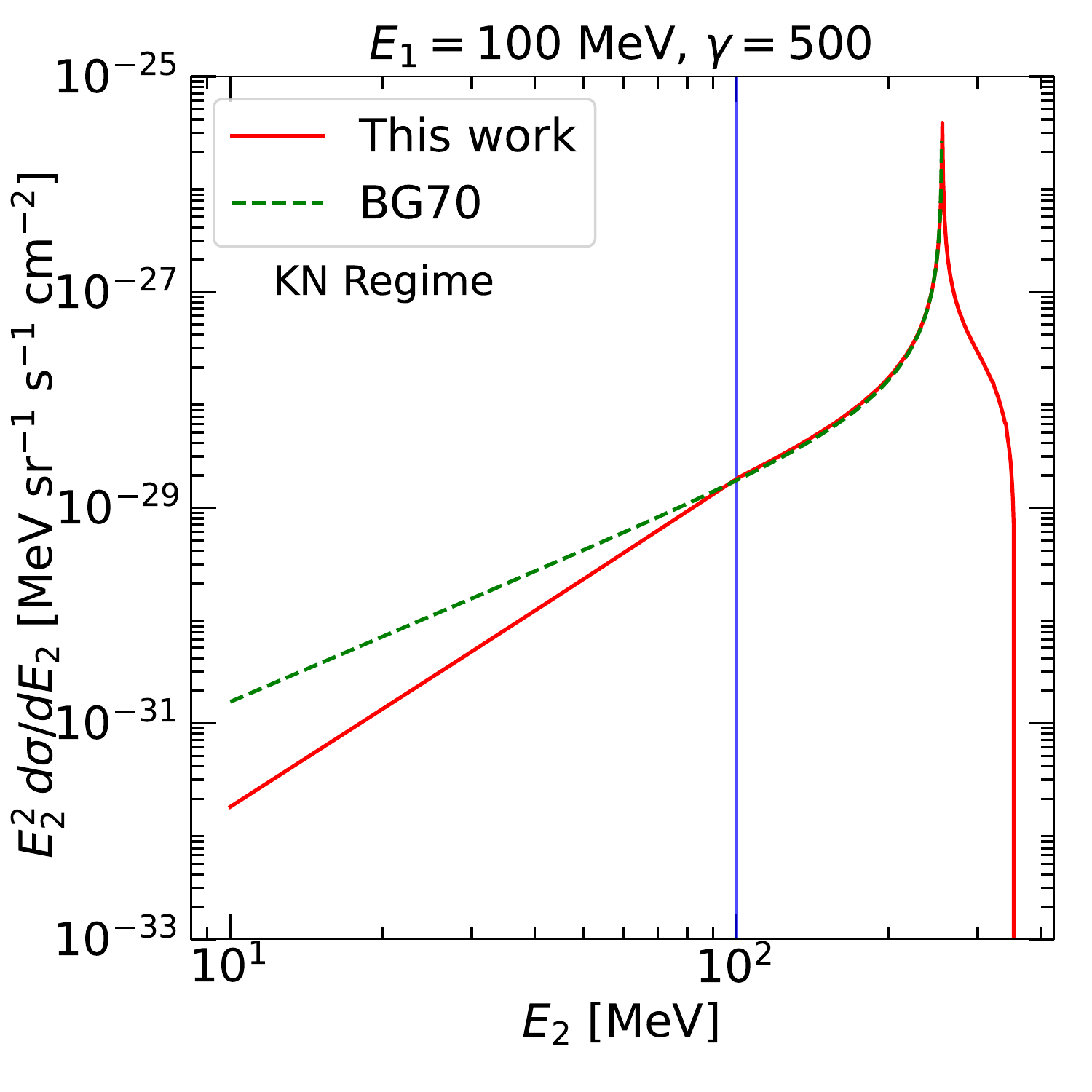}
        \label{fig:soft_KN_larger_gamma}
    \end{subfigure}
    \caption{Similar to Fig.~\ref{fig:Thomson_single_photon_electron}, but for the KN regime, $E'_1 \gtrsim m_e$, $E_1 \lesssim E_e $~(Sec. \ref{sec:Eph_l_Ee}). Left: $E_1 = 0.5$ MeV and $\gamma = 1.5$, which corresponds to mildly-relativistic scattering; Right: $E_1 = 100$ MeV and $\gamma = 500$, which corresponds to ultrarelativistic scattering.}
    \label{fig:soft_KN_single_photon_electron}
\end{figure*}

\par
Our exact formalism is expected to agree well with BG70 in the ultrarelativistic limit.  However, in the mildly-relativistic limit, when the $\gamma$ value is smaller, the approximation $E_{2, \rm max} \sim 4\gamma^2 E_1$ from head-on scattering between source electron and photon no longer holds. In this case, the upscattered photon energy
\begin{equation}\label{eq:E2_thomson_regime}
    E_2 \simeq \frac{1 - \beta\cos\theta_e}{1 - \beta\cos\theta_1}E_1
\end{equation}
can be obtained by taking the limit $E_1 \ll E_e$ in Eq.~\ref{eq:explicit_E2} or by considering the relation $E_1' = E_2'$ and express $E_1', E_2'$ in observer frame quantities. The exact $E_{2, \rm max}$ is then yielded by maximizing Eq.~\ref{eq:E2_thomson_regime} with respect to $\theta_{e}$ and $\theta_{1}$.

\par
The left panel in Fig.~\ref{fig:Thomson_single_photon_electron} plots the differential cross section against $E_2$ for the mildly-relativistic case of $\gamma = 1.5$ electrons scattering with source photons at $10^{-6}$ MeV. The scattering falls into the Thomson regime as $E_1' \approx 3\times 10^{-6}$ MeV. The blue line marks the value of $E_1$ and separates the spectrum into upscattering ($E_1 < E_2$) and downscattering ($E_1 > E_2$) regions. In our exact calculation, the cutoff energy is lower than BG70, and the differential cross section shifts to the left. Our differential cross section also differs from BG70 in the down-scattering region, which we discuss in detail later in Sec.~\ref{sec:final_remark}.

The right panel of Fig.~\ref{fig:Thomson_single_photon_electron} shows the same plot but with $\gamma = 100$ for source electrons to depict the ultrarelativistic IC scattering. The scattering is still in the Thomson regime as $E_1' \sim 2 \times 10^{-3}$\,MeV $\ll m_e$. As expected, in the upscattering domain $E_2 > E_1$, our result agrees well with BG70, and have produced the same $E_{2, \rm max} = 4 \gamma^2 E_1 \approx 4 \times 10^{-2}$ MeV.
In addition, the total Thomson cross section $\sigma_T = 8\pi/3~r_e^2$ can be recovered by integrating the area under $d\sigma/dE_2$ in Fig.~\ref{fig:Thomson_single_photon_electron} over $E_2$. Therefore, it validates our formalism on the differential cross section in the ultrarelativistic limit. 

\subsubsection{KN regime ($E_1 \lesssim E_e $, $E'_1 \gtrsim m_e$)}\label{sec:Eph_l_Ee}

For larger values of $E_1$ that approaches $E_e$, the ERF target photon energy $E_1'$ can easily surpass the rest mass of the electron $m_e$; this corresponds to the KN regime. In this regime, BG70 approximated the maximum scattered photon energy $E_{2, \rm max}$ to be:
\begin{equation}\label{eq:KN_E2max}
    E_{2, \rm max} \approx \frac{4\gamma^2 E_1}{1 + 4\gamma \frac{E_1}{m_e}}\, ,
\end{equation}
which follows from applying the large $\gamma$, head-on scattering with the photon scattered backward approximation ($\theta_e = \pi, \theta_1 = 0$) to the conservation of energy Eq.~\ref{eq:explicit_E2}.  When $E_1' \gg m_e$, the term $4\gamma E_1/m_e$ in the denominator of Eq.~\ref{eq:KN_E2max} dominates and $E_{2, \rm max} \simeq E_e$.  Using the general differential cross section, we instead find that $E_{2, \rm max} = (\gamma - 1)E_e + E_1$, which is simply the case when the electron transfer all its kinetic energy to the photon and is valid for any values of $\gamma$.

\par
The left panel in Fig.~\ref{fig:soft_KN_single_photon_electron} shows the differential cross section of $\gamma = 1.5$ electrons scattering with 0.5 MeV target photons. As $E_1' \approx 2\gamma E_1 = 1.5$ MeV, which implies that photons undergo Compton scattering in ERF and a large portion of the energy is transferred to the electron. This corresponds to the KN regime.

\begin{figure*}[t!]
    \centering
    \begin{subfigure}
        \centering
        \includegraphics[width=\columnwidth]{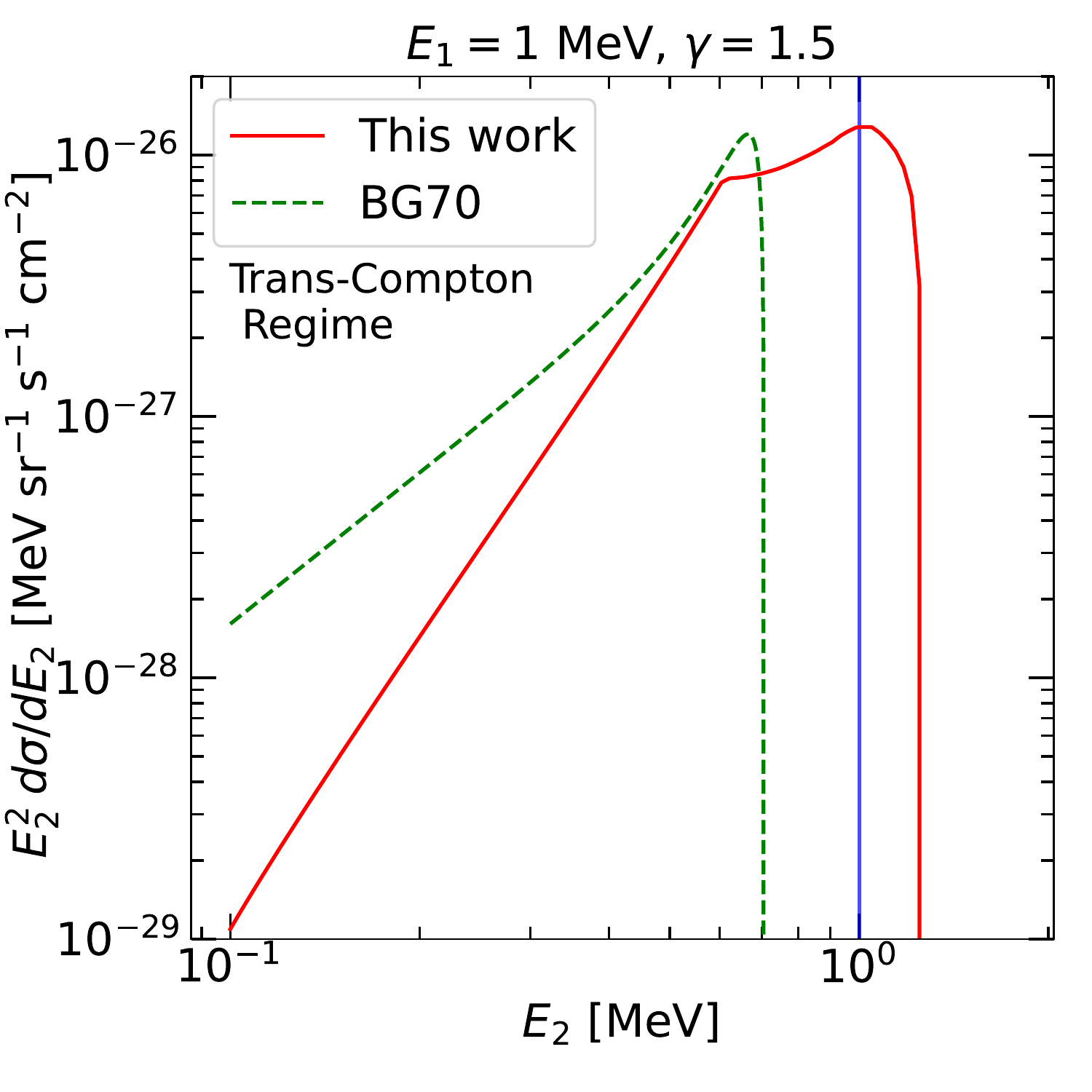}
        \label{fig:hard_KN_small_gamma}
    \end{subfigure}
    \begin{subfigure}
        \centering
        \includegraphics[width=\columnwidth]{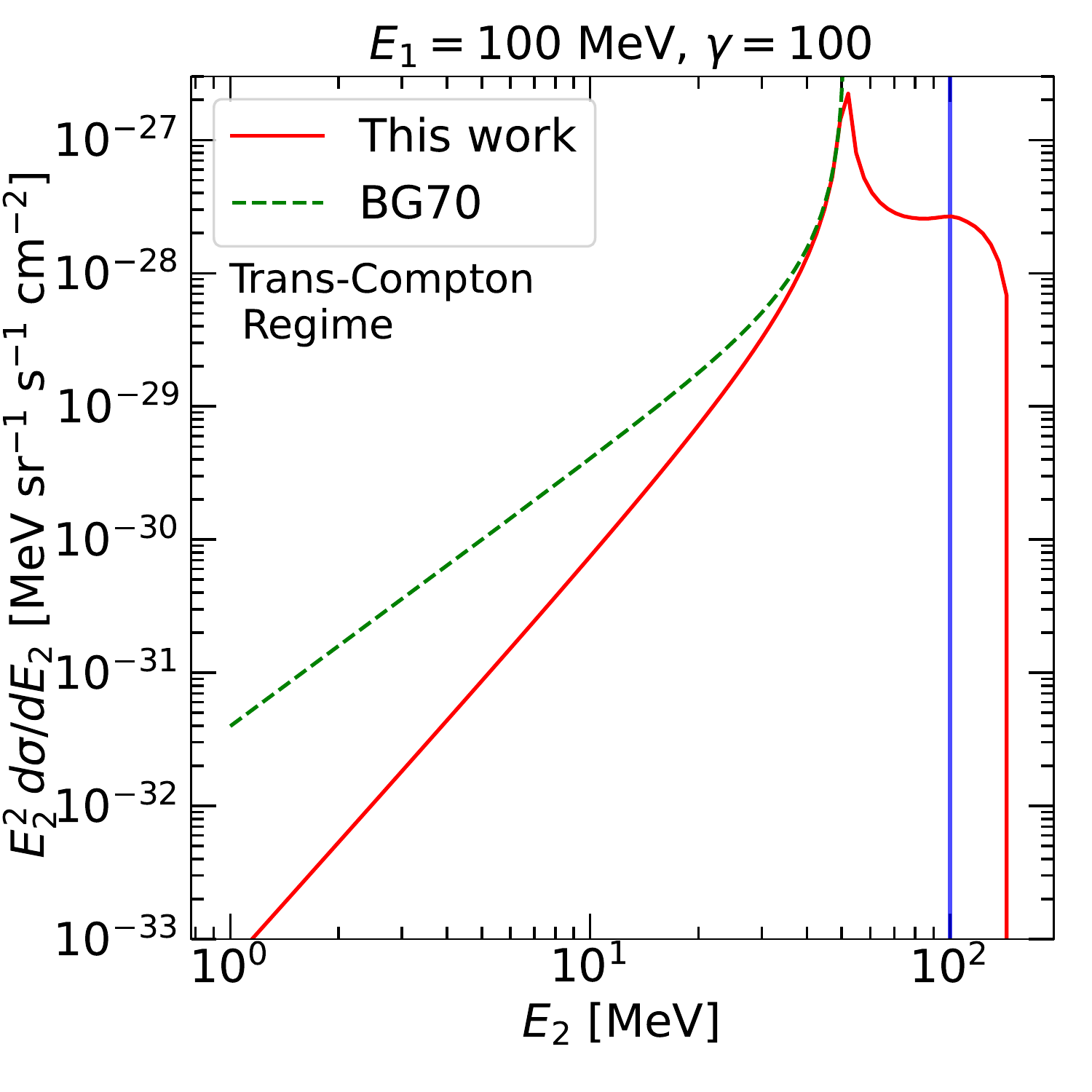}
        \label{fig:hard_KN_larger_gamma}
    \end{subfigure}
    \caption{Similar to Fig.~\ref{fig:Thomson_single_photon_electron}, but for the trans-Compton regime, $E_{1} > E_e$~(Sec. \ref{sec:Eph_g_Ee}). Left: $E_1 = 1$ MeV, $\gamma = 1.5$, corresponds to mildly-relativistic scattering; Right: $E_1 = 100$ MeV, $\gamma = 100$, corresponds to ultrarelativistic scattering.}
    \label{fig:hard_KN_single_photon_electron}
\end{figure*}

The right panel in Fig.~\ref{fig:soft_KN_single_photon_electron} shows the ultrarelativistic scattering $\gamma = 500$ electron and 100 MeV target photons, which is in the regime of $E_1' \approx 2\gamma E_1 \gg m_e$. As expected, our results agree with BG70 in the $E_1 < E_2  < E_e$ energy range. BG70's formula, however, does not work above $E_e$, while our results extend correctly to the true maximum at $E_{2} = (\gamma - 1)E_e + E_1$. 
We note that for large $\gamma$ cases, the differential cross section strongly peaks at the electron energy $E_e$. This can be clearly seen in this plot (as well as in the right panel of Fig.~\ref{fig:hard_KN_single_photon_electron}.)

\par

\subsubsection{Trans-Compton regime ($E_1 > E_e$)}\label{sec:Eph_g_Ee}

When $E_1 > E_e$, that is, when the target photon energy is greater than the electron energy, the scattering in ERF entirely falls in the KN regime. But in this case, the target photon mostly experiences energy loss, similar to the case of Compton scattering. Although the case of $E_1 > E_e$ was not considered in BG70, we compare to their formula for completeness. Our formalism works for all energy range up to $E_{2, \rm max} = E_1 + (\gamma - 1) E_e$.

The left panel in Fig.~\ref{fig:hard_KN_single_photon_electron} shows the differential cross section of $\gamma = 1.5$ electrons scattering with 1 MeV target photons, which corresponds to a mildly-relativistic scattering. The ultrarelativistic case ($\gamma = 100$, $E_1 = 100$ MeV) is illustrated in the right panel of Fig.~\ref{fig:hard_KN_single_photon_electron}. The large deviation of BG70's differential cross section from ours implies the breakdown of ultrarelativistic approximation in this kinematic regime. 

\subsubsection{Down-scattering cases ($E_2 < E_1$)}\label{sec:final_remark}

In all the cases discussed above, our general formalism includes the effect of down scattering (when $E_2<E_1$), which is not included in BG70. From Fig.~\ref{fig:Thomson_single_photon_electron} to Fig.~\ref{fig:hard_KN_single_photon_electron}, we note that the correct differential cross sections always decline more rapidly than BG70 in the down scattering regime. In addition, we also correctly calculate the minimum scattered photon energy $E_{2, \rm min}$ using Eq.~\ref{eq:explicit_E2}, which corresponds to a photon and an electron initially travelling in the same direction with the photon scattered backward ($\theta_e = 0,~\theta_1 = \pi$). In contrast, there is no $E_{2, \rm min}$ from BG70.

\section{Case studies}\label{sec:spectrum_photon_electron}
We consider two simple cases of high-energy IC emission to show that our formalism can correctly reproduce the results in ultrarelativistic limits, and find the regime where the ultrarelativistic assumption would break down. 

\subsection{Solar Inverse Compton Emission}\label{sec:solaric}
Solar IC emission are produced when cosmic-ray electrons up scatter solar photons~\cite{moskalenko:2006_solar_photons_modulation_and_neutrino_physics, orlando:2007_extened_IC_gamma_ray_from_sun, Orlando:2007_gamma_rays_from_halos_around_sun, orlando:2008_gamma_ray_from_solar_halo_and_disk_egret}. 
The anisotropic IC scattering cross section in MS~\cite{Moskalenko:2000_anisotropic_IC_in_galaxy} (with ultrarelativistic approximations) was adopted to produce the IC emission package \stellarics~\cite{orlando:2013_stellarics_stellar_solar_iC_emission_package}. In this section, we compare our exactly formalism with the latest \stellarics~calculation on solar gamma ray in Ref.~\cite{Orlando:2021_stellarics_sun_from_kev_to_tev}.

\par
\begin{figure}[t!]
    \centering
    \includegraphics[width=\columnwidth]{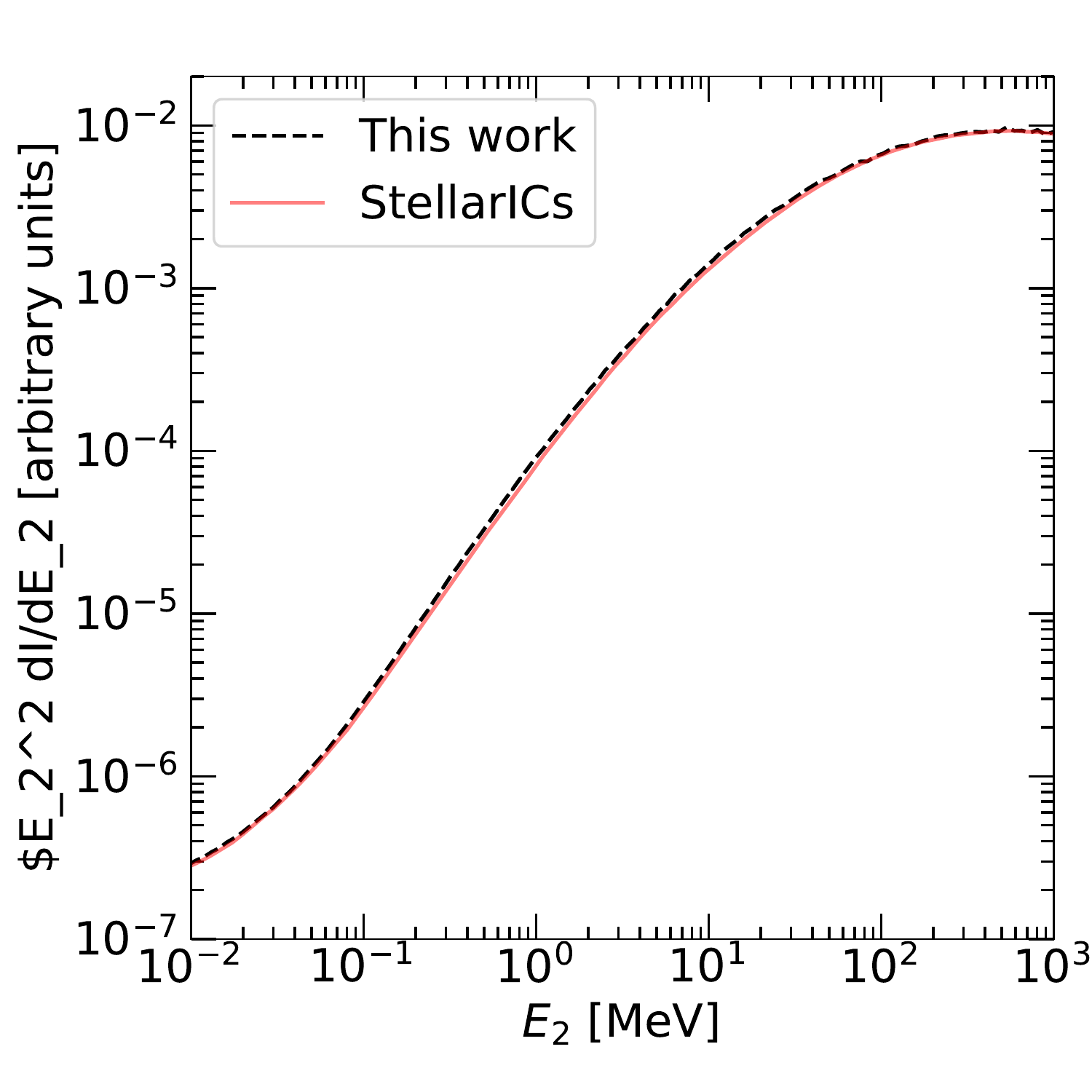}
    \caption{The LOS solar IC intensity at $0.3^\circ$ away from the centre of the sun. The red spectrum labelled with '\stellarics' refers to the solid blue curve in ~\cite{Orlando:2021_stellarics_sun_from_kev_to_tev} Fig. 4. The intensity spectrum in this work was computed using Eq.~\ref{eq:master_equation} with the same solar photon and cosmic electron spectrum in ~\cite{AMS:2014}. In particular, a thermal spectrum from $10^{-7}$ to $10^{-5}$ MeV was used for the photon field. An electron spectrum from $10^{1}$ to 10 $^{5}$ MeV that follows the curve fitted with AMS02 data in Fig. 3 of ~\cite{Orlando:2021_stellarics_sun_from_kev_to_tev}, was used.}
    \label{fig:solaric_spectrum}
\end{figure}

Fig.~\ref{fig:solaric_spectrum} shows the LOS solar IC spectral intensity from the general electron-photon scattering formalism and \stellarics. The computation of our spectrum follows the master equation Eq.~\ref{eq:master_equation} with an observation angle $\theta_0 = 0.3^\circ$. We employ the same $n_{ph}, f_{ph}, n_{e}, f_{e}$ as in~\stellarics. Specifically, the differential number density of target photons follows a black-body spectrum at 5770 K, with a spherical and uniform distribution. The differential number density of source electrons is inferred from the curve fitted with AMS02 data in the Fig. 3 of ~\cite{Orlando:2021_stellarics_sun_from_kev_to_tev} and the distribution is isotropic. From Fig.~\ref{fig:solaric_spectrum}, our spectrum agrees well with \stellarics~in the range of $E_2 = 10^{-2} - 10^{3}$ MeV. The correction from releasing the ultrarelativistic approximation in the differential cross section cannot be seen in the figure, since electrons with $\gamma = 10^2 - 10^4$ are responsible for scattering solar photons to the range of hard X-rays and gamma rays. The general formula thus reduces to ultrarelativistic approximation in this regime and converge to \stellarics's.

\begin{figure}[t!]
    \centering
    \includegraphics[width=\columnwidth]{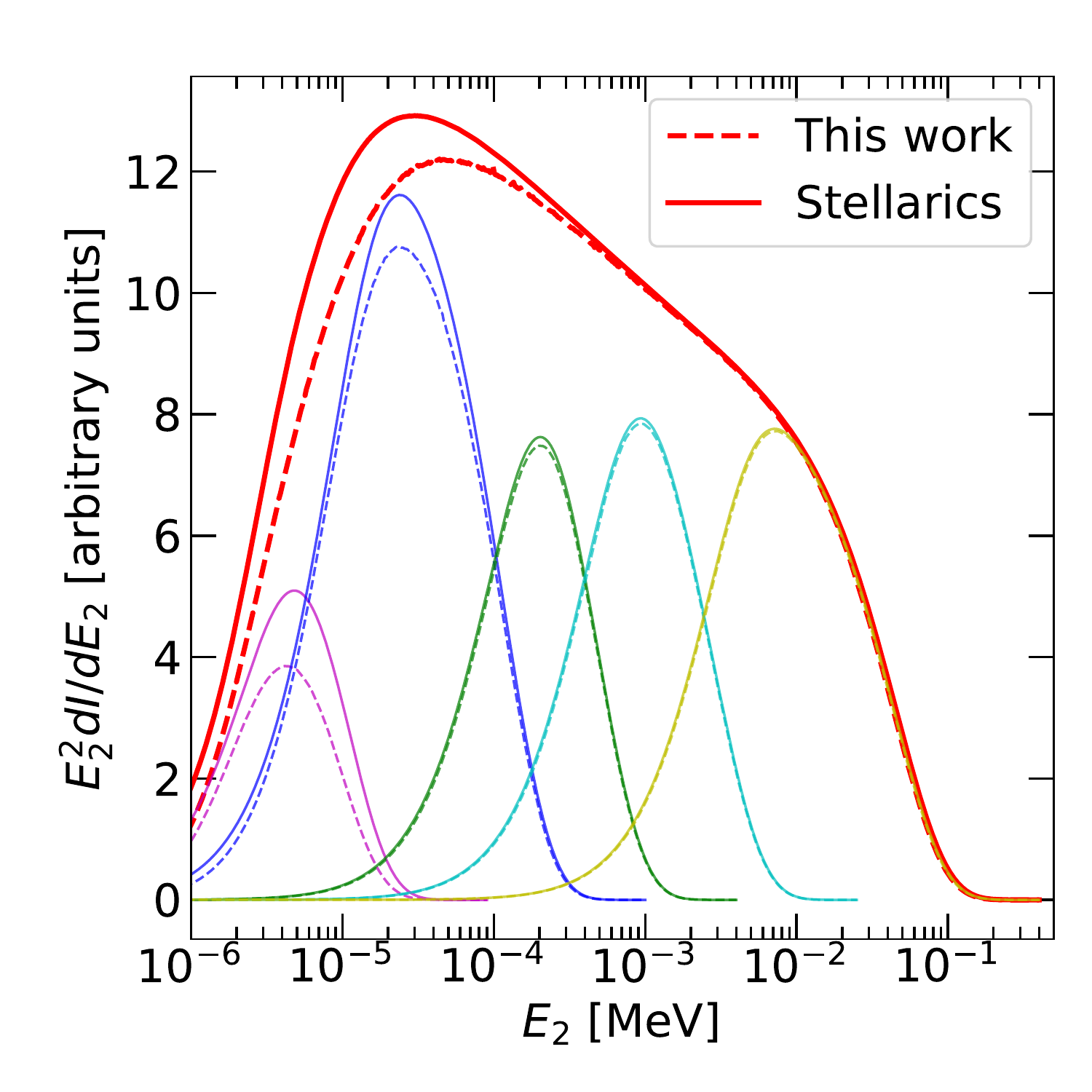}
    \caption{A decomposition of LOS solar IC spectrum. The decomposed spectra with $\gamma = 1 - 1.5$, $\gamma = 1.5 - 5$, $\gamma = 5 - 10$, $\gamma = 10 - 25$ and $\gamma = 25-100$ are given by the magenta, blue, green, cyan, and yellow curves, respectively. Solid lines represent the~\stellarics~results and our results (general formalism) are shown in dashed lines. Linear vertical scale is used to highlight the mildly-relativistic correction.}
    \label{fig:stellarics_decomposition}
\end{figure}

For smaller values of $E_2$, we expect to see some discrepancies between our exact calculation and the ultrarelativistic approximations.  For illustration, we extend the electron spectrum to lower energies, adopting a $E_e^{-3.2}$ power law between $\gamma = 1$ to 100. 

Fig.~\ref{fig:stellarics_decomposition} shows the full energy range for the anisotropic solar IC scattering. It is clear that the solar emission above $E_2 \approx 10^{-3}$ MeV is indeed dominated by larger $\gamma > 10$. The ultrarelativistic approximation holds in this regime and the two emission spectra coincide. 
When $E_2 < 10^{-3}$ MeV, our calculation deviates from \stellarics. From the spectrum decomposition, we see the deviation is due to mildly-relativistic correction from general formalism. The deviation is the largest in the interval $\gamma = 1.5 - 5$. The overall correction to the total emission is a reduced intensity for $E_2 < 5\times10^{-3}$ MeV.

\subsection{Synchrotron Self-Compton}\label{sec:ssc}
Synchrotron photons are emitted when energetic electrons gyrate along strong magnetic fields. These synchrotron photons can also undergo IC scattering with the gyrating electrons, resulting high-energy gamma-ray emission. This synchrotron self-Compton (SSC) mechanism have been used to model gamma-ray emissions from various sources, such as blazars and relativistic jets \cite{Potter:2012jet3, Inoue:2019jet1, Banik:2019jet2}.

In the SSC mechanism, the target photon energy is higher than the solar IC case.  We therefore consider a simplified SSC model to see the effect of relaxing the ultrarelativistic assumption. 

We consider a relativistic jet, where the Lorentz factor of the jet is 500 with a $0.5^\circ$ observation angle from the jet direction.
The resulted boost from the jet frame (JF) to the observer frame is about 50. 
Both the electron and photon spectra are taken to be isotropic in JF so that we can compare the result with BG70. In the JF (quantities denoted with the ``JF'' subscript), we consider the electron spectrum to be a power law $E_{e, \text{JF}}^{-3}$ between $\gamma_\text{JF} = 1$ to $ 10^{4}$. The photon spectrum is taken to be a $E_{1, \text{JF}}^{-2}$ power law between $10^{-3}$ MeV to $10^{3}$ MeV.

\begin{figure}[t!]
    \centering
    \includegraphics[width=\columnwidth]{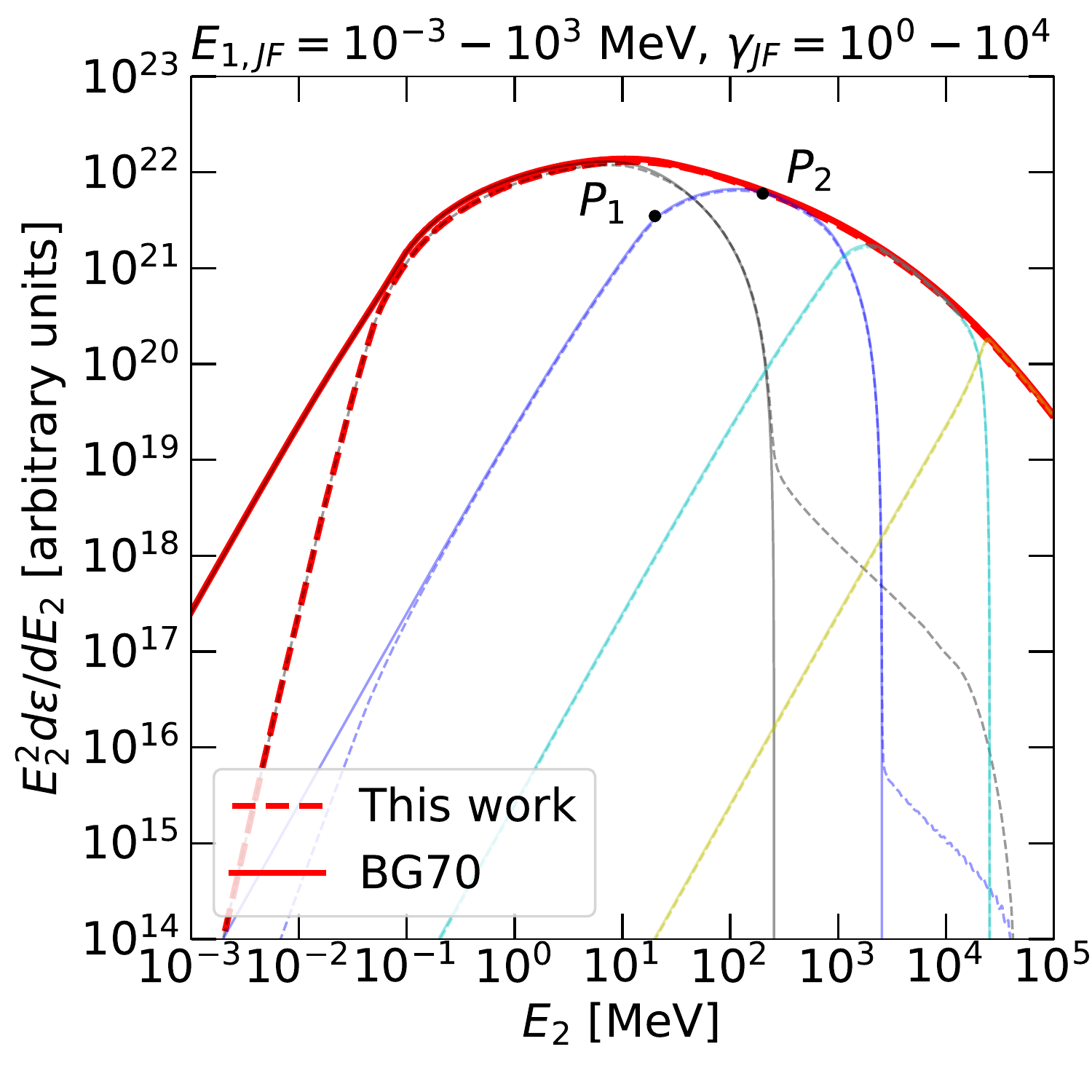}
    \caption{The SSC emission spectrum at observation angle $0.5^\circ$ away from the jet with $E_{1, \text{JF}}$ from $10^{-3}$ to $10^{3}$ MeV and $\gamma_\text{JF}$ from 1 to $10^{4}$. Both of target photons and electrons are assumed to be isotropic in JF. The emission is plotted in scattered photon energy in observer frame $E_{2}$. Red line gives the total emission. The individual contributions from $\gamma_\text{JF} = 1 - 10^1, 10^1 -10^2, 10^2 - 10^3$, and $10^3 - 10^4$ are presented in the black, blue, cyan, and yellow lines, respectively. The transition points $P_1$ and $P_2$ for the blue line are highlighted. See the text for details about the decomposition spectrum and transition points. }
    \label{fig:SSC_spectrum}
\end{figure}

Fig.~\ref{fig:SSC_spectrum} shows the SSC emission spectrum $E_2^2 d\epsilon/dE_2$ in the observer frame with the above configuration. The red dashed line represents the total emission from our calculation. For comparison, results obtained from BG70 is shown in solid lines. Our results agrees well to that obtained with BG70 at high photon energies, but large deviations start to appear at low energy. 

For $E_{2, \text{JF}}$ smaller than the minimum photon energy, $10^{-3}$ MeV ($E_{2}$ $\approx 0.05$ MeV, in Fig.~\ref{fig:SSC_spectrum}), these are dominated by down-scattered photons produced in the Thomson regime. 
In this case, BG70's formula corresponds to a flat differential cross section, and thus results in a $E_2^2$ behaviour in the $E_2^2 d\epsilon/dE_2$ plot. 
As discussed in Sec.~\ref{sec:final_remark}, our results deviates considerably from the ultrarelativistic approximation.   

To further illustrate the physics of the IC scattering, consider the blue line in Fig.~\ref{fig:SSC_spectrum}, which is the spectrum produced by electrons with $\gamma_\text{JF}$ between $10^{1}$ and $10^{2}$.  For $E_{2} > 0.05$ MeV ($E_{2, \text{JF}} > 10^{-3}$), the spectrum enters a region where all scattering geometries are accessible by Thomson scattering, thus resulting in the $E_2^2$ behaviour.  The $E_2^2$ trend continues until reaching the point $P_1$, which corresponds to $E_{2, \text{JF}} = 4\gamma_\text{JF,m}^2 E_{1, \text{JF,m}}$, where both $\gamma_\text{JF,m}$ and $E_{1, \text{JF,m}}$ are the lower cutoff of the source photon and electron spectrum.   Above $P_1$, the spectrum then softens mainly due to the spectral shape of source photons.  Finally, above the point $P_2$, the scattering transits into the KN regime, and the spectrum steepens further mainly due to the spectral shape of source electrons.  We note that the transition into the KN regime can be rapid in the case of large $\gamma$ values.  E.g., for the case of the the yellow line in Fig.~\ref{fig:SSC_spectrum}.

Finally, as mentioned in Sec.~\ref{sec:Eph_l_Ee}, our formalism works correctly for cases with $E_2 > E_{e, \rm{max}}$ when there are appreciable energy in the target photon. These contribution appears as the 'bump' features in Fig.~\ref{fig:SSC_spectrum}. In these cases, the extra contributions are overwhelmed by those from higher energy electrons. 

\section{Conclusions and Outlook}\label{sec:Conclusions}

In this work, we consider the scattering between energetic electrons and photons, and study the differential cross section for the outgoing photon spectrum. This process is frequently used in high-energy astrophysics for the production of IC photon emission.  

We demonstrate that for the general case (without ultrarelativistic approximation) of anisotropic photons scattering with isotropic electrons, the differential cross section can be written as Eq.~\ref{eq:explicit_averaged_cross_section}, which can be easily integrated numerically by finding the analytic solutions to the scattering geometry, given by Eq.~\ref{eq:theta_e_equation}. 

Focusing on isotropic photon and isotropic electron scatterings, we compare our formalism with that from BG70, which considered the ultrarelativistic approximation. We find that the scattering can be divided into three regimes, the Thomson regime, the KN regime, and the Trans-Compton regime.  In general, we find that there would be deviations to the BG70 formula in the downscattering limit ($E_{2}< E_{1}$), as well as when the target photon energy becomes comparable to the electron energy.  

To validate our formalism in the ultrarelativistic limit and show that it is numerically stable, we consider two cases of high-energy IC emission.  The first is solar IC emission produced by anisotropic solar photon scattering with isotropic cosmic-ray electrons. We find that our results agree well with that from \stellarics~\cite{Orlando:2021_stellarics_sun_from_kev_to_tev}.  Due to solar photons being relatively low in energy, corrections from the exact formalism only appear below keV photon energy. 
The second case we consider is SSC emission, where isotropic synchrotron photons can scatter with isotropic electrons in astrophysical jets. Due to the synchrotron photons being comparatively energetic, the corrections from downscattering can be significant in the low energy part of the SSC.  At high energies, our results agree well with that produced by the BG70 formula.  

In these case studies, although we only consider scattering between isotropic source electrons with isotropic photons~(SSC) or anisotropic photons~(solar IC), it is straightforward to generalise the calculations to include different angular distributions for photons and electrons in Eq.~\ref{eq:ang_avg_cross_section}. For example, for the case of external Compton emission in astrophysical jets~\cite{Dermer:1995_EC1, Kelner:2014_anisotropic_electrons}, photons emission are produced by jet electrons scattering with photons from CMB, accretion disk, or dusty torus~\cite{Finke:2016_EC2}. In these cases, both photon and electron distribution can be anisotropic, and thus a more general formalism like ours is required.  

In this work, we have focused much of our discussion in the gamma-ray regime, where IC emission is typically considered.  This is in part to show that our formalism can be reduced to the well established works in the ultrarelativistic regime. In general, we expect that a general formalism like this is required whenever mildly relativistic electrons are involved, or when the target photon energy is not small compared to electron energy. 

Finally, through Eq.~\ref{eq:theta_e_equation}, we have obtained an exact solution to the scattering angle geometry. This allows us to obtain the photon polarization caused by anisotropic scattering, which we will discuss in detail in a followup work~\cite{Lai:polarization}.  In the literature, results of photon polarization caused by IC scattering are somewhat inconsistent~\cite{krawczynski:2012_polarization_of_IC_in_blazar}.  This work forms the basis for a numerical framework to obtain the polarization spectrum. This is especially timely, given that there are recent and future  X-ray and gamma-ray telescopes that are capable of detecting photon polarizations~\cite{IXPE:2016, eXTP:2016, ASTROGAM:2021}. 

\section{Acknowledgement}
We thank Ming-Chung Chu for helpful discussions. This work makes use of~\stellarics, an open source code that is available from Ref.~\cite{Orlando:2021_stellarics_sun_from_kev_to_tev}. This project is supported by a grant from the Research Grant Council of the Hong Kong Special Administrative Region, China (Project No. 24302721).

\bibliography{bib.bib}

\appendix
\section{Relating positive determinant to the kinematic constraint of $E_2$}\label{app:positive_determinant}
\par
A non-negative determinant in the quadratic equation Eq.~\ref{eq:solve_thetae} $\Delta = A^2 + B^2 - C^2 \geq 0$ secures real solutions and sets the kinematic constraint on the possible range of $E_2$. To illustrate this, rewrite the determinant into a quadratic equation of $E_2$:
\begin{eqnarray}\label{eq:quadratic_eqation_in_E2}
        &&K_1E^2_2 + K_2 E_2 + K_3 \geq 0 \\
        K_1 &=& \beta^2(\cos^2\theta + \sin^2\theta\cos^2\phi_e) - \left[\frac{E_1(1 - \cos\theta)}{E_e} + 1\right]^2 \nonumber \\
        K_2 &=& 2E_1\left[\frac{E_1(1 - \cos\theta)}{E_e} + 1 - \beta\cos\theta\right] \nonumber \\
        K_3 &=& (\beta^2 - 1)E^2_1 \nonumber .
\end{eqnarray}

Eq.~\ref{eq:quadratic_eqation_in_E2} has another determinant $\Delta' = K_2^2 - 4K_1K_3$ which is positive definite for a physical solution. In the non-relativistic KN regime, $\beta \rightarrow 0$, $E_1 \sim m_e \approx E_e$, We have $K_1 < 0$ such that $E_2$ is bounded the roots of Eq.~\ref{eq:quadratic_eqation_in_E2}. We also have $\Delta' \approx 0$, so:
\begin{eqnarray}\label{eq:NR_KN_E2_limit}
        \frac{E_1}{\frac{E_1}{m_e}(1-\cos\theta) + 1} &\leq& E_2 \leq \frac{E_1}{\frac{E_1}{m_e}(1-\cos\theta) + 1} \\
        \implies E_2 &=& \frac{E_1}{\frac{E_1}{m_e}(1-\cos\theta) + 1} = E_\text{Compton} \nonumber \, ,
\end{eqnarray}
as expected.

\par
In the Thomson regime, $E'_1 \ll m_e$ and $E_2$ is again bounded by the roots as $K_1 < 0$. We further consider ultrarelativistic limit such that $\gamma \ll 1$, both maximum and minimum scattered photon energy $E_2$ occurs at $\cos\theta = -1$ scattering in this limit. Manipulation on $\Delta'$ yields:
\begin{align}\label{eq:determinant_approx}
    \begin{split}
        K^2_2 - 4K_1K_3 &= (1 + \beta)^2 - (1 - \beta^2)^2\\
        &= (1 + \beta)^2[1 - (1 - \beta)^2]\\
        &\approx (1 + \beta)^2\left(1 - \frac{1}{4\gamma^4}\right), 
    \end{split}
\end{align}
where the relativistic approximation $\beta \approx 1 - 1/(2\gamma^2)$ is used in last line. With Eq.~\ref{eq:determinant_approx} and additional calculations, the conditional $\Delta \geq 0$ becomes:
\begin{eqnarray}\label{eq:UR_TH_E2_limit}
        \frac{(1-(1-1/(8\gamma^4))}{1 - \beta}E_1 \leq &E_2& \leq \frac{1+(1-1/(8\gamma^4))}{1 - \beta}E_1 \nonumber \\
        \frac{1}{4\gamma^2}E_1 \leq &E_2& \leq 4\gamma^2 E_1.
\end{eqnarray}

Again, it agrees with the limits mentioned in~\cite{Rybicki:1986}. Therefore, a real solution of Eq.~\ref{eq:theta_e_equation} implies a physical solution for the scattering.
\end{document}